# Biologically generated turbulent energy cascade in shear flow depends on tensor geometry

Xinyu Si[1] and Lei Fang[1,*]

[1]Department of Civil and Environmental Engineering, University of Pittsburgh, Pittsburgh, PA 15261, USA
[*]Corresponding author: lei.fang@pitt.edu

October 5, 2023

**Abstract**

It has been proposed that biologically generated turbulence plays an important role in material transport and ocean mixing. Both experimental and numerical studies have reported evidence of the non-negligible mixing by moderate Reynolds number swimmers in quiescent water, such as zooplankton, especially at aggregation scales. However, the interaction between biologically generated agitation and the background flow as a key factor in biologically generated turbulence that could reshape our previous knowledge of biologically generated turbulence, has long been ignored. Here we show that the geometry between the biologically generated agitation and the background hydrodynamic shear can determine both the intensity and direction of biologically generated turbulent energy cascade. Measuring the migration of a centimeter-scale swimmer-as represented by the brine shrimp *Artemia salina*-in a shear flow and verifying through an analogue experiment with an artificial jet revealed that different geometries between the biologically generated agitation and the background shear can result in spectral energy transferring toward larger or smaller scales, which consequently intensifies or attenuates the large scale hydrodynamic shear. Our results suggest that the long ignored geometry between the biologically generated agitation and the background flow field is an important factor that should be taken into consideration in future studies of biologically generated turbulence.

Biologically generated (biogenic) turbulence has been considered to have great potential in scalar transport and ocean mixing. Earlier back-of-envelope estimations have proposed that the marine biosphere can mix the ocean as effectively as winds and tides [4]. It has also been estimated that the dissipation rate of schooling-animal-induced turbulence was comparable to the turbulence resulting from major storms [16]. Even though this idea received challenges in later years for the low mixing efficiency [34, 18], it has been proved in both laboratory experiments [15, 36] and numerical simulations [28] that centimeter-scale swimmers with moderate Reynolds numbers can have a non-negligible effect on ocean mixing, especially at aggregation scales [15].

Although the biogenic turbulence from micro-scale bacteria to dekameter-scale ocean mammals [16, 35, 5, 29] has been extensively investigated, the biogenic turbulence of the swimmers was mostly examined independent of background flows. However, most aquatic organisms live in bodies of water that are nearly always dynamically flowing. For instance, diurnal vertical migrations of zooplankton [32, 14] repetitively penetrate horizontal hydrodynamic shear layers driven by wind [3] and internal waves [10] in lakes and oceans. When we consider the coupling between the swimmers' agitation and the background flow, from the first principle of the spectral energy flux [7], one can realize that the geometry between the biogenic agitation and the background flow plays a key role, and different geometric configurations could lead to qualitatively different and even opposite consequences.

Here we propose and experimentally verify a new mechanism of biogenic turbulence in background hydrodynamic shears that has long been omitted in biogenic turbulence studies. This mechanism involves the geometric alignment between a small-scale turbulent stress tensor arising from centimeter-scale swimmers' agitation and a large-scale rate of strain tensor arising from background hydrodynamic shears, such as the ubiquitous wind-driven hydrodynamic shear in oceans and lakes. We show that the different geometric configurations between these two tensors can result in turbulent kinetic energy transferring to either smaller or larger scales, which can further intensify or attenuate large-scale shear, causing completely opposite mixing consequences.

Before we introduce the new mechanism in biogenic turbulence, we quickly review the concept of tensor geometry in classical turbulence [30, 21, 7]. As with any energy flux, turbulent cascades can be interpreted as the result of the action of a stress against a rate of strain. Thus, it is clear that the relative geometry between the turbulent stress and the large-scale rate of strain is important for determining the direction and magnitude of the spectral energy flux. This picture of spectral energy flux can be numerically resolved via filter-space technique (FST) [13, 22, 6, 26, 21, 7, 1]. FST is conceptually similar to the approach taken in Large Eddy Simulation (LES) [24], with the primary difference being that filtering is undertaken as a preceding step before solving the equations in LES and as an a posteriori step in an FST. The core of FST is the application of a spectral low-pass filter to the velocity field. If we set the cutoff length scale to be $L$, then all the velocity variations on scales smaller than $L$ are suppressed, while all the large-scale variations are preserved. Taking the inner product between the filtered Navier-Stokes equation and the filtered velocity field $u_i^{(L)}$, we can obtain the evolution equation for the filtered kinetic energy $E^{(L)} = (1/2) u_i^{(L)} u_i^{(L)}$. This equation includes a new term which represents the energy transfer through length scale $L$ with the form [20, 21]

$$Q^{(L)} = -\tau_{ij}^{(L)} S_{ij}^{(L)}, \tag{1}$$

where $\tau_{ij}^{(L)} = (u_i u_j)^{(L)} - u_i^{(L)} u_j^{(L)}$ is a turbulent stress tensor and $S_{ij}^{(L)} = (1/2)(\partial u_i^{(L)}/\partial x_j + \partial u_j^{(L)}/\partial x_i)$ is the rate of strain tensor of the filtered velocity field. Eqn. 1 is essentially the inner product of the filtered stress tensor and the rate of strain tensor. When $Q^{(L)} < 0$, it indicates inverse energy flux toward larger length scales, while $Q^{(L)} > 0$ indicates forward energy flux toward smaller length scales.

In two dimensions, $Q^{(L)}$ can be reexpressed as a function that depends on the geometric alignment between the eigenframes of the rate of stain and the stress as [21, 7]

$$Q^{(L)} = -2\lambda_\tau^{(L)} \lambda_S^{(L)} \cos(2\theta^{(L)}), \tag{2}$$

where $\lambda_S^{(L)}$ and $\lambda_\tau^{(L)}$ are the largest eigenvalues of the deviatoric rate of strain and stress tensors, and $\theta^{(L)}$ is the angle between the corresponding (extensional) eigenvectors. It is then clear that the alignment of the stress and the rate of strain tensor can determine not only the magnitude but also the direction of the energy flux. Previous researchers have suggested treating $\cos(2\theta^{(L)})$ as the efficiency of the energy flux between scales [7]. When $\theta^{(L)} < \pi/4$, energy flux to larger scales generate inverse energy flux; when $\theta^{(L)} > \pi/4$, energy flux to smaller scales generate forward energy flux. There is no net energy flux if $\theta^{(L)} = \pi/4$.

In the LES context, the stress tensor can be further decomposed into Leonard stress, cross stress, and subgrid-scale Reynolds stress based on the type of triad interactions [19]. Previous work has shown that the energy flux due to the subgrid-scale Reynolds stress ($\tau_S^{(L)} = ((u_i - u_i^{(L)})(u_j - u_j^{(L)}))^{(L)}$) contributes dominantly to the net spectral energy flux $Q^{(L)}$ [20]. The Leonard stress and the cross stress, on the other hand, play a negligible role in the net energy flux between scales [33]. In addition, because of the limited interrogation area, practically, the Leonard stress and the cross stress, which should have trivial contributions for the net energy flux $Q^{(L)}$, can lead to significant contamination for $Q^{(L)}$ in the filtering process near domain boundaries due to edge padding. Therefore, we use the energy flux corresponding to the subgrid-scale Reynolds stress ($Q_s = -\tau_{S,ij}^{(L)} S_{ij}^{(L)}$) instead of $Q^{(L)}$ in the following analysis (see Supplementary information). It is worth noting that other decomposition methods also exist and one of the most commonly used methods was proposed by Germano [12, 13], which gives all three components in a Galilean-invariant way. The $Q_s$ term using both two methods gives qualitatively the same result (see Supplementary information).

Enlightened by this mechanistic picture of tensor geometry in Navier-Stokes turbulent flows, we consider the additional "biogenic stress", i.e., additional stress due to swimmers' agitations, on top of the background hydrodynamic shear. At low Reynolds numbers, a large-scale laminar hydrodynamic shear should have negligible spectral energy flux. Typical centimeter-scale swimmers in oceans and lakes, such as zooplankton, generate jets behind them while propelling themselves (see Fig. 1b) [15, 36, 28]. A simple analysis can reveal that, effectively, the extensional eigenvector of the stress tensor due to the agitation of a swimmer coincides with its swimming direction (see Supplementary information). If a centimeter-scale zooplankton is introduced in a large-scale hydrodynamic shear, it is obvious that the relative orientation between the zooplankton's swimming direction and the large-scale shear layer's extensional rate of strain eigenvector plays a key role in determining the effect of the biogenic turbulence because the biogenic agitation can couple with the large scale shear layer to generate either forward or backward energy flux. From the first principle, we attain a new mechanism of biogenic turbulence of centimeter-scale swimmers in background hydrodynamic shear. We call the angle between the extensional eigenvector of the hydrodynamic shear's rate of strain tensor and the swimming direction of the centimeter-scale swimmer $\theta_s$ (see Fig. 1c). Then, we can reasonably predict that when $\theta_s < \pi/4$, energy fluxes to larger scales; when $\theta_s > \pi/4$, energy fluxes to smaller scales. Particularly, when $\theta_s = \pi/4$, the spectral energy transfer should be the same as a swimmer swimming in quiescent water. In addition, we hypothesize that if swimmers collectively and repetitively swim through a shear layer with $\theta_s < \pi/4$, the shear layer will be enhanced due to the constant inverse energy flux. On the other hand, if the collective and repetitive migration of swimmers has $\theta_s > \pi/4$, we hypothesize that the forward energy flux will attenuate the shear layer because part of the energy contained in the large-scale shear layer is constantly fed to smaller scales. We predict that the shear strength will be intact if $\theta_s \approx \pi/4$.

We experimentally examined this new biogenic turbulence mechanism and tested the hypothetical consequences by conducting laboratory experiments with an electromagnetically driven shear flow and a representative swimmer: *A. salina*. *A. salina* was chosen because it is a typical centimeter-scale zooplankton with a swimming mode that is similar to various saltwater and freshwater zooplankton [36]. In our experiments, the directions of extensional eigenvectors for both the large-scale strain rate of the shear and the small-scale stress of the jet behind the swimmer can be easily determined (see Fig. 1c).

We put a thin layer of 14% by mass saltwater above two columns of magnets with opposite magnetic poles (Fig. 1a). The distance between the two magnet columns was 10 cm. This distance was long enough to prevent the two magnet columns from interacting so that the magnetic field was self-closed near each column of magnets. Two extra lateral boundaries were settled above the magnet columns to separate the middle shear region from the strong flow directly above the magnet columns. By passing a DC current through the water, we could generate a stable shear flow at the middle of the setup using the resulting Lorentz body force. We seeded fluorescent tracers (with diameters between 106 and 125 $\mu$m ) into the flow. A 7 cm by 7 cm area (mapping to a 1024 pixel by 1024 pixel camera sensor) at the middle of the shear could be highly resolved through a particle tracking velocimetry (PTV) algorithm that captures about 9,000 particles each frame at a frame rate of 60 frames per second [27]. The flow was stable and stream-wise (y direction) uniform. Figure 2a shows the characterization of the measured shear flow. Within the 7 cm width, the flow velocity increases almost linearly from –1.2 cm/s to 1.2 cm/s. The extensional eigenvector of the strain rate tensor exhibited a uniform orientation towards the $\pi/4$ direction along the x-axis throughout the full image domain (Fig. 1c and Fig. 2a). The root-mean-square velocity in the full image domain was 0.77 cm/s.

*A. salina* was introduced into the shear flow system and was allowed to move freely. Video sections that capture a single swimmer passing the interrogation area were selected and classified into three categories: $\theta_s < \pi/4$, $\theta_s > \pi/4$, and $\theta_s \approx \pi/4$. We waited long enough between the video sections to let the fluctuations in laminar shear decay to a negligible level. Additionally, two more sets of videos were collected for comparison. One set was taken with the pure shear flow without *A. salina* introduced and the other was taken with only *A. salina* in the field and no background shear flow.

Using FST, the spectral energy flux was calculated for each set with a cutoff length *L* from scales smaller than the body length of *A. salina* to scales that are comparable to the shear flow. The results perfectly confirmed our prediction from the first principle about the relationship between the spectral energy flux directions and $\theta_s$. As is shown in figure 2b, for length scales between the swimmer-body length and the shear flow, cases with $\theta_s < \pi/4$ show an inverse energy flux that is much stronger than the no shear cases and the pure shear case. Cases with $\theta_s > \pi/4$ show a strong forward energy flux. Cases with $\theta_s \approx 0$ show an energy flux similar to cases without background shear and the energy flux intensity soon converges to zero when the cutoff length scale increases above the swimmer-body length. These results allow us to make two conclusions. First, the geometry between *A. salina*'s swimming direction and the background shear does play a significant role in determining the result of biogenic turbulence. Second, its small length scale notwithstanding, agitation induced by the *A. salina* can couple with the flow at much larger scales because we observe strong net spectral energy fluxes at scales that are much larger than the individual length when $\theta_s < \pi/4$ and $\theta_s > \pi/4$. For example, Figure 2c shows the spatial distribution of spectral energy flux through a length scale of 4 cm, which is noticeably larger than the body length of an *A. salina*. We observe that the dominant contribution for the spectral energy flux is around the swimmer, indicating that the coupling between the agitation from the swimmer and background shear is indeed dominant for the net energy flux.

Note that at the length scale comparable to an *A. salina*'s body, all cases with swimmers show an inverse energy flux. This is because, at small scales, the local shear is strongly deformed by *A. salina*'s agitation and, therefore, the original direction of the extensional eigenvector of the rate of strain tensor no longer holds and the two-dimensional flow naturally generates inverse energy flux simply due to its dimension [17].

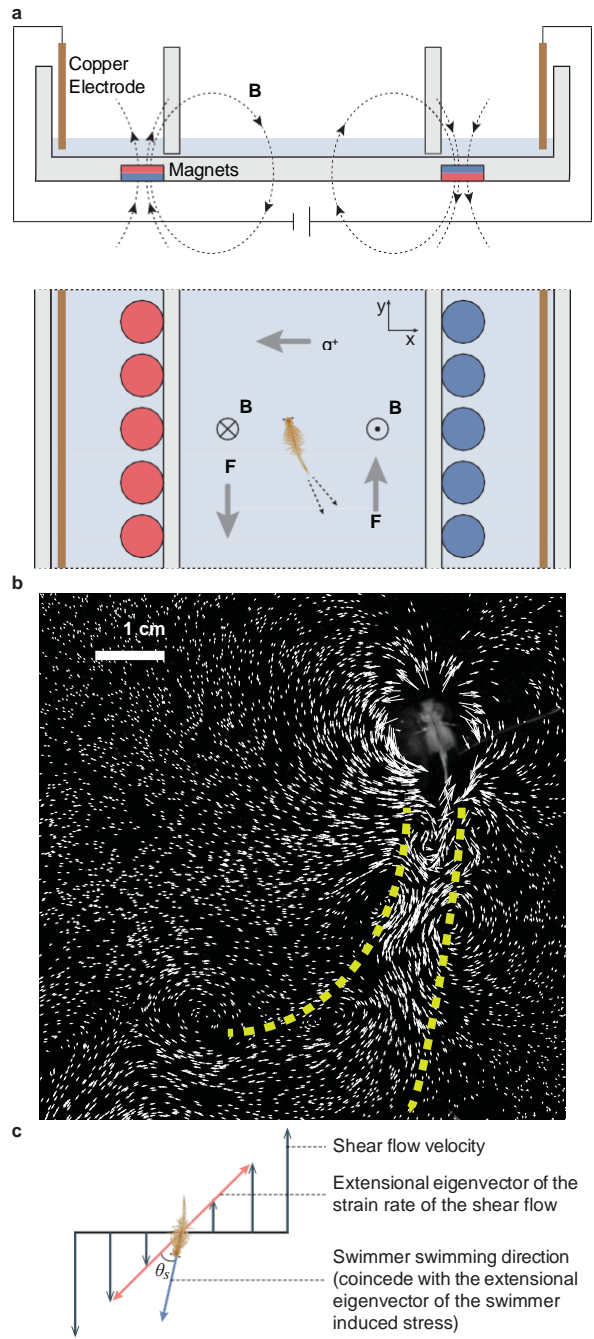

Figure 1: **Experimental setup and visualization of biogenic flow. a**, schematic of the electromagnetically driven shear flow system. The front view image gives the direction of the magnetic field. The resulting in-plane Lorentz body force F drives the hydrodynamic shear. **b**, measured velocity field near a single *A. salina* in quiescent water. Except for the near-body field, the major flow induced by a swimming *A. salina* is a jet opposite to its swimming direction. The yellow dashed lines mark the rough profile of the jet. **c**, schematic of tensor geometry for biogenic turbulence in a hydrody5namic shear.

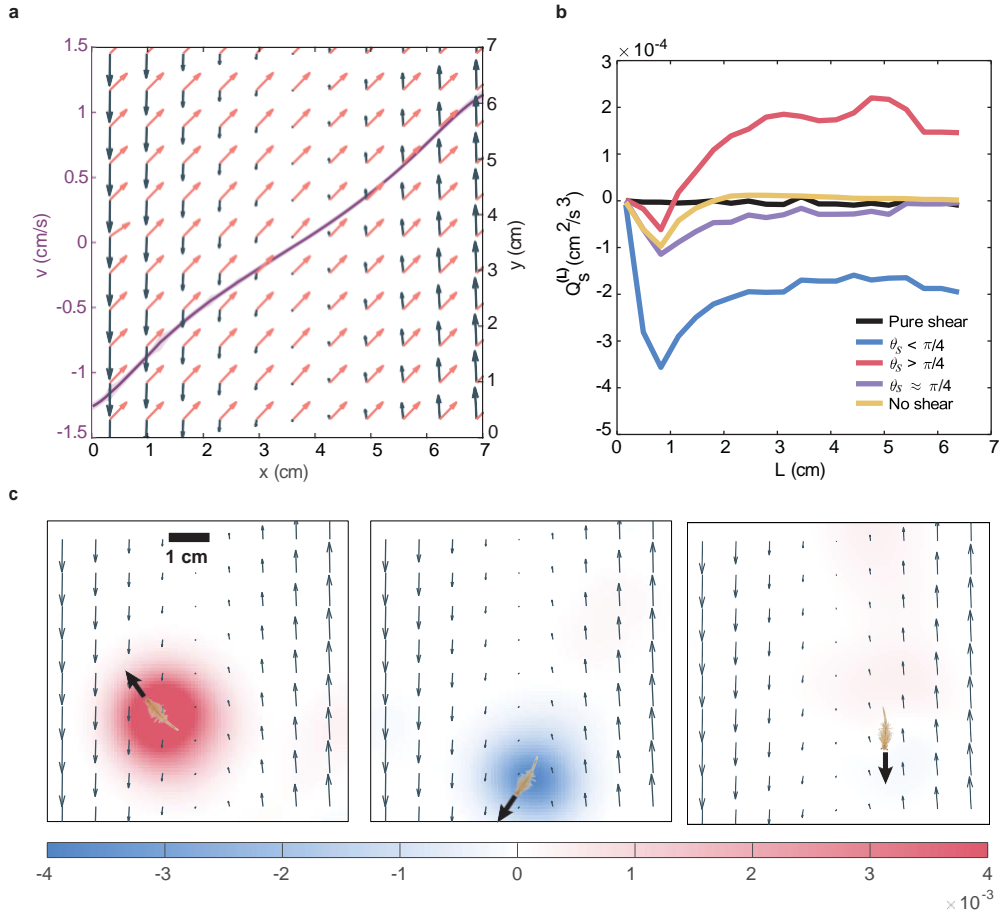

Figure 2: **Shear characterization and spectral energy flux of biogenic turbulence.**
**a**, characterization of the shear flow generated by the electromagnetically driven shear flow system. The dark gray vectors show the down-sampled mean velocity field. The pink vectors show the local extensional eigenvector directions of the mean velocity field. The purple curve gives the v component velocity profile along the x direction. The v component is averaged over the y axis and over 3000 frames. The one standard deviation range around the mean velocity profile is given in the narrow shaded area but too small to observe. **b**, the net spectral energy flux with different tensor geometries between *A. salina* and shear. **c**, a single frame of $Q_s$ with filter length scale $L$ = 4 cm. Gray vectors show the down-sampled mean shear flow. The *A. salina* schematic and the black vector in each panel together indicate the location and the swimming direction of the shrimp.

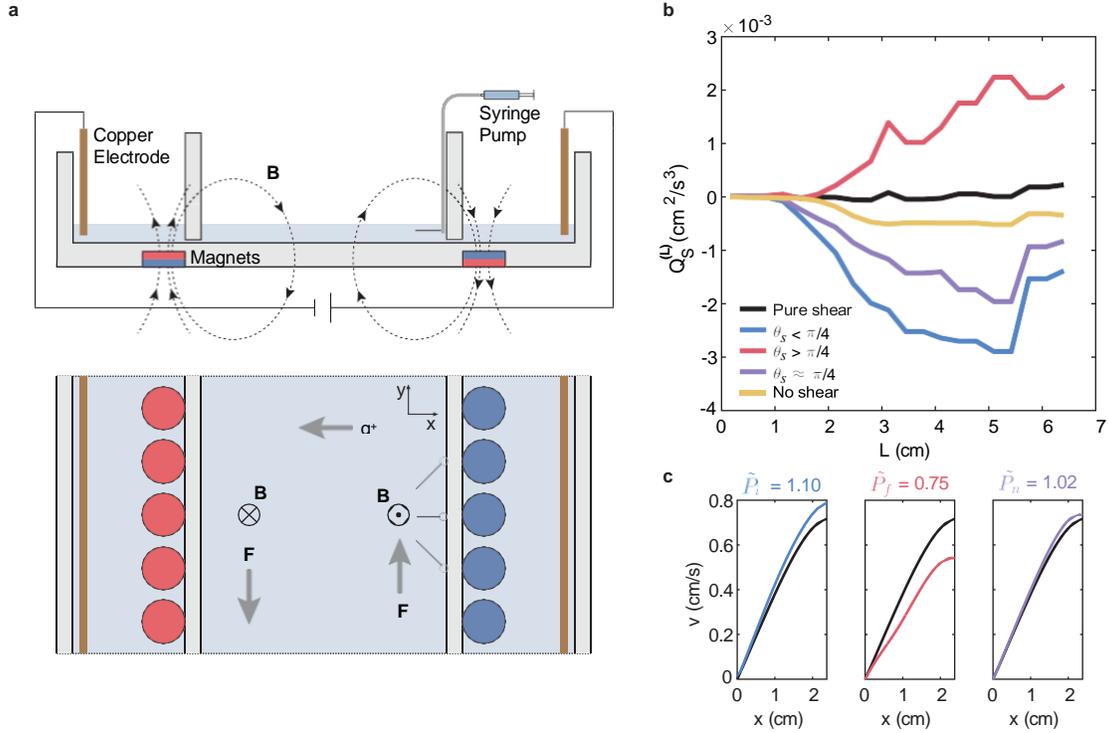

Figure 3: **Experimental setup and results of the analogue experiment. a**, schematic of the setup for the analogue experiment. The dispensing needles are fixed to the inner boundary and connected to a syringe pump. **b**, spectral energy flux for cases with different tensor geometry configurations. **c**, profiles of mean v component of velocity along the x direction. Line colors are consistent with that of subfigure b. The title for each subpanel gives the normalized mean velocity gradient. The original shear is intensified in the inverse flux case (the first subpanel) while it is attenuated in the forward flux case (the second subpanel). The neutral case (the third subpanel) has a mean velocity gradient similar to the original shear.

The agitation introduced by the *A. salina* interacted with the shear, fluxing energy either to larger or smaller scales. As a result, we hypothesize that the inverse/forward energy flux during a collective and repetitive swimmer migration through a hydrodynamic shear can enhance/attenuate the large-scale shear. When $\theta_s > \pi/4$ and forward net energy flux dominates, the large-scale shear is expected to be attenuated because part of the large-scale energy needs to be fed into smaller scales. On the other hand, for $\theta_s < \pi/4$ cases, the inverse energy flux is expected to enhance the large-scale shear flow, strengthening the velocity gradient in the large-scale shear. This is analogous to the energy condensation in two-dimensional flow [37, 9], where energy piles up at the largest scale allowed by the system when the scale at which the energy is dissipated exceeds the domain size. The condensed largest-scale energy enhances the largest-scale structure in physical space. The difference between our case and energy condensation is that an energy dissipation mechanism exists in our system. The energy removal mechanism is linear friction at a large scale, and the dissipation rate is linearly proportional to the kinetic energy at a large scale [8], i.e., the shear scale. Due to the enhanced inverse energy flux, the linear dissipation mechanism requires a larger kinetic energy pile-up at a large scale for energy balance, leading to a stronger shear.

Since the energy of a single *A. salina* is too small to create measurable effects for the large-scale shear and a swarm-scale accurate, angle-specific control for *A. salina* is not feasible in laboratories, to examine the effect of stronger biogenic turbulence, we designed an analogue experiment with jets of fixed angles (Fig. 3a). As is shown in Fig. 1b, the major flow induced by a single swimming *A. salina* is a jet opposite to its swimming direction. It was also shown that large swarms could generate swarm-scale coherent jets that resulted in most of the mixing effect [28]. Therefore, an artificial jet is a good proxy for a stronger jet created by collectively moving swarms. We use jets created by a dispensing needle to mimic the jets created by a small group of swimmers. Based on the electromagnetically driven shear setup, we fixed three thin dispensing needles with 0.96 mm inner diameter to the right inner boundary, with each needle holding a different angle corresponding to one tensor geometry configuration, i.e., $\theta_s = 0$, $\theta_s = \pi/4$ and $\theta_s = \pi/2$. As a proxy of a swarm swimming collectively in the same direction, the needles were connected to a syringe pump that gave a stable discharge at 6 mL/min, which created stable jets in different directions. To mimic the small swarm passing and leaving the shear layer, the syringe pump was turned on and off intermittently at a mean interval of 14 s. For each angle, six video sections recorded from the start to the full development of the jet were selected and analyzed. Since we were mainly focused on the interaction between the additional stress due to swimmers' agitations ("biogenic stress") and large scale rate of strain tensor, we focused on a 2.5 cm by 3.3 cm space near the needles where the local spectral energy flux was the strongest. Our results were robust for a range of reasonable domain sizes near the dispensing needles.

In comparison with the spectral energy flux for a single *A. salina* (Fig. 2b), the stronger jet flow led to the energy flux increasing by one order of magnitude (Fig. 3b). Unlike in the experiment with *A. salina*, there was no clear length scale for the jet flow. Therefore, the energy flux curves did not show an inverse flux at small length scales. The energy flux for the $\theta_s = \pi/4$ case was stronger than that for the no-shear case, even though it was still weaker than the $\theta_s < \pi/4$ case. This is because the stronger jet can expand into a much larger length scale, disturbing the local shear, and hence the original rate of strain's extensional eigenvector direction no longer holds. In two dimensions, a flow naturally has net inverse energy flux [17].

To examine the effect of tensor geometry on modifying large-scale flow field, we calculated the profile of the v component velocity field along the x direction (Fig. 3c). The v component was averaged over the y direction and normalized to zero at x = 0. In addition to calculating the velocity profiles of the shear layers, we also calculated the mean velocity gradient $P = \langle dv/dx \rangle$, where the angle bracket indicates the spatiotemporal mean. For different geometric configurations, the mean velocity gradient $P$ was normalized using the mean velocity gradient of the pure shear flow, denoted as $\tilde{P}$. Consistent with our hypothesis, the shear flow was intensified in the inverse flux case ($\tilde{P}_i = 1.10$) and attenuated in the forward flux case ($\tilde{P}_f = 0.75$). The neutral case ($\theta \approx \pi/4$) had a velocity gradient similar to that of the pure shear flow ($\tilde{P}_n = 1.02$), and the velocity profiles were also very similar. The modification of the shear layer is highly nontrivial given the fact that the amount of energy injected by the needles was orders of magnitudes smaller than that of background shear because the root-mean-square velocity was not altered by the additional jet (see Supplementary information). As the stress becomes stronger, the intensification of the large scale shear under inverse flux condition may trigger Kelvin–Helmholtz instabilities that will consequently further facilitate the grow of biologically generated turbulence.

Results with a single *A. salina* and the analogue experiments consistently showed that the geometry between swimmer-generated agitation and background flow is a key factor in biogenic mixing, a fact that has long been overlooked in biogenic turbulence studies. Our results shed new light on biogenic turbulence with complex background flows. Since diurnal vertical migrations often extend from the surface to depths up to 600 meters, swimmers have enough chance to encounter and penetrate horizontal hydrodynamic shears frequently [2]. Moreover, because our conclusion regarding the critical role that geometric configuration plays in biogenic mixing is from the first principle of energy flux, the importance of tensor geometry in biogenic turbulence can be extended to various background flow structures besides hydrodynamic shears, such as ocean eddies and jets.

Furthermore, this new mechanism in biogenic turbulence could reshape previous conclusions about how a biosphere's physical feedback affects regional and global ocean flows. For instance, the strengthened or attenuated shear due to biogenic turbulence may have profound feedback on local transport and mixing [31, 23], thereby affecting regional productivity and growth [25, 11].

# Materials and Methods
## 1.1 Electromagnetically driven shear system and particle tracking velocimetry

The two-dimensional shear flow system was built using acrylic with lateral dimensions of 35 cm by 35 cm. Beneath the bottom floor, we put two columns of magnets with opposite polarities that covered the full expanse of the setup. Each magnet (neodymium grade N52) had an outer diameter of 1.27 cm and thickness of 0.64 cm with a maximum magnetic flux density of 1.5 T at the magnetic surface. Within each column, the magnets were closely packed with a center to center space of 1.53 cm so that the magnetic field could distribute uniformly beside each side of the magnetic columns. The distance between the center of the two magnet columns was 10 cm. This distance was large enough to prevent the two magnet columns from interacting and hence the magnetic field beside each column of magnets was self-closed.

We loaded a thin layer (0.5 cm) of 14 % by mass NaCl solution with a density of 1.1 g/cm$^2$ and viscosity of $1.25 \times 10^{-2}$ cm$^2$/s above the bottom floor. A pair of copper electrodes were placed at opposite sides of the setup. By passing a DC current (0.4 A) through the conducting layer, we could drive a quasi-two-dimensional flow using the resulting Lorentz body force. To prevent the strong flow directly above each magnet column from interacting with the middle shear flow, we added two inner boundaries at the center of the setup that created a 35 cm by 8.5 cm space, within which the Lorentz body force formed a well-controlled hydrodynamic shear.

To track the flow, we seeded the fluid with green fluorescent polyethylene particles (Cospheric) with a density of 1.025 g/cm$^3$ and diameters ranging from 106 to 125 $\mu$m. The Stokes number was low enough (of order $10^{-3}$) to accurately track the flow. A small amount of surfactant was added to prevent a clustering effect of particles due to surface tension, and the surface tension was too small to affect the movement of the tracers, which was confirmed by a negligible tracer speed with no driven force. We used a machine vision camera (Basler, acA2040-90$\mu$m) to record a 7 cm by 7 cm region at the center of the setup with 1024 pixels by 1024 pixels resolution. Only the images that captured a single swimmer passing through the recorded region were selected and analyzed. Swimmers in the images were pre-masked and then the images were processed by a particle tracking velocimetry (PTV) algorithm [27]. About 9000 particles can be recorded and tracked with a frame rate of 60 frames per second. The combination of the particle seeding density and frame rate allowed a highly spatiotemporally resolved Eulerian velocity field to be measured. We further interpolated the measured fluid properties, such as velocity, onto a regular Eulerian grid through cubic interpolation with grid size $\Delta x$ = 10 pixels (0.07 cm).

## 1.2 Swimmer

Adult sized *A. salina* were obtained (Northeast Brine Shrimp) and cultured in 3 % by mass NaCl solution at least 24 hours before conducting the experiment. During the experiment, the *A. salina* was introduced into the middle of the shear region and was allowed to swim freely. Four LED light panels were mounted around the experimental setup to create a uniform light environment so that the movement of swimmers would show no directional bias.

## 1.3 Analogue experiment

The analogue experiment setup was modified based on the original electromagnetically driven shear system. On the right inner boundary, we fixed three thin dispensing needles with inner diameters of 0.96 mm that were parallel with the bottom floor. The tips of all three needles had the same distance to the right inner boundary. Each needle held a different angle with respect to the hydrodynamic shear ($\pi/4$, $\pi/2$, $3\pi/4$ with the positive y direction, respectively). The three jet directions mimicked $\theta_s = 0$, $\theta_s = \pi/4$ and $\theta_s = \pi/2$ cases, respectively. During the measurement of each different angle, the corresponding needle was connected to a syringe pump (KD Scientific, Legato 100) that could give a stable discharge at 6 mL/min. The syringe pump was turned on and off intermittently at a mean time interval of 14 s. For each case, six video sections that had recorded from the start to the full development of the jet were selected for PTV analysis.

# Supplementary Material

Supplementary material is available at PNAS Nexus online.

# Funding

This work is supported by the U.S. National Science Foundation under Grant No. CMMI-2143807.

# Author contributions statement

L.F. conceived the original idea and supervised the project. X.S. ran the experiments and analyzed the data. Both authors wrote the paper.

# Data availability

Data files and MATLAB scripts to reproduce all the figures in this article and the SI files are available on github via https://github.com/FangResearchGroup/Swimmers.git.